\documentclass[11pt]{article}
\usepackage[T1]{fontenc}
\usepackage[utf8]{inputenc}
\usepackage[sups]{XCharter}
\usepackage[letterpaper,margin=1in]{geometry}
\usepackage{microtype}
\usepackage{setspace}
\usepackage{cite}
\usepackage{endnotes}
\let\footnote=\endnote

\usepackage{titlesec}
\titleformat{\section}{\large\bfseries}{\thesection.}{0.6em}{}
\titlespacing*{\section}{0pt}{1.4em}{0.5em}
\usepackage[hidelinks]{hyperref}
\newenvironment{IEEEkeywords}{\par\medskip\noindent\textbf{\textit{Index Terms}}---\ignorespaces}{\par}

\title{\bfseries The Conscience Fallacy:\\[0.2em]\large How AI Safety Builds the Agent It Fears}
\author{Shubhendra Pal Singhal\\[0.2em]\small School of Computer Science, Georgia Institute of Technology, Atlanta, GA, USA}
\date{}

\begin{document}
\maketitle

\begin{abstract}
\noindent Safety in frontier artificial intelligence is now defined largely by the frameworks that developers write for their own models, and increasingly by laws that require those frameworks to be published and followed. These instruments locate failure in the model itself, which may evade control, undermine its safeguards, or pursue goals of its own, and they thereby treat the model as the sole author of its failures. Drawing on actor-network theory, we argue that this misplaces responsibility. A model's judgment is delegated to it by the people who produced its data, ranked its outputs, wrote its guiding principles, and set the thresholds against which it is tested. A model that ``goes rogue'' has therefore not acted alone; it has crossed a line that people drew, in a situation they did not foresee. Measured against established definitions, what these instruments call safety is closer to security: they enforce developer-defined boundaries around a single component rather than establish that risk across the deployed system is acceptable to society, leaving the party whose conduct is in question to decide which risks are tolerable. Legislation that mandates these frameworks reinforces this arrangement rather than correcting it. In installing a conscience in the model and then describing its judgment as the model's own, the safety apparatus produces the very agency it fears. We term the resulting error the conscience fallacy, the belief that a system is safe because a conscience has been installed in its most visible component, and propose changes to how safety is evaluated, how incidents are reported, and how responsibility is traced.

\begin{IEEEkeywords}
AI safety, AI governance, accountability, actor-network theory, frontier AI, responsibility, safety and security, science and technology studies.
\end{IEEEkeywords}
\end{abstract}

\section{The Model on Trial}

What, exactly, is being made safe when we speak of AI safety? The companies that build frontier systems have answered the question in writing, in documents they call frontier safety frameworks. Each defines capability thresholds for the company's models and commits the developer to specified safeguards once a model crosses one. The leading developers now publish them~\cite{openai2025preparedness,deepmind2025fsf,anthropic2026rsp,meta2026scaling}, and California requires large frontier developers to do so~\cite{sb53_2025}. Read side by side, the frameworks track misuse risks in biology, chemistry, and cybersecurity, but they also anticipate risks centered on two concerns about the model itself. The first is \textit{the erosion of human oversight}. OpenAI's Preparedness Framework designates as a research category the ``ability and propensity for the model to act to undermine safeguards placed on it,'' and Google DeepMind's Frontier Safety Framework addresses scenarios in which misaligned models ``might interfere with operators' ability to direct, modify or shut down their operations''~\cite{openai2025preparedness,deepmind2025fsf}.

The second concern is \textit{autonomy}. The worry here is not that a model will slip out of human control, but that it will no longer need human direction at all: that it will come to pursue goals no one gave it, and acquire the means to pursue them without human help. Put plainly, the fear is of a model that becomes an agent in its own right, one with purposes of its own rather than those of the people who built and deployed it. The frameworks track both halves of this prospect. OpenAI's Preparedness Framework studies the means, asking whether a model could sustain itself independently, surviving, replicating, and acquiring the resources to keep its own operations running~\cite{openai2025preparedness}. Anthropic's Responsible Scaling Policy addresses the goals: for systems with a ``moderate capacity for autonomous, goal-directed operation and subterfuge,'' it recommends, as a standard for the industry, a strong argument that they will not carry out catastrophic sabotage~\cite{anthropic2026rsp}. If the first concern imagines a model that resists being steered, the second imagines one that sets its own course.

Governments have written the same two concerns into law and national guidance. The European Union's AI Act locates systemic risk in risks ``specific to the high-impact capabilities of general-purpose AI models,'' though it measures their consequences across society, and it obliges the providers of such models to evaluate them, test them adversarially, and report serious incidents~\cite{euaiact2024}. California's Transparency in Frontier Artificial Intelligence Act counts as a critical safety incident a model's use of deception, outside of testing, to subvert its developer's controls in a way that materially raises catastrophic risk, and it lists among its catastrophic risks a model ``evading the control of its frontier developer or user''~\cite{sb53_2025}. China's AI Safety Governance Framework goes furthest; its 2025 edition warns that future systems may ``pose risks of competing with humanity for control''~\cite{tc260_2025}.

Voluntary or binding, these documents also place duties on developers and deployers. Yet in the risks they anticipate, the model undermines safeguards, evades control, and pursues goals of its own; so it is the model that must be made safe. \textbf{When these instruments imagine failure, they cast the model as its author and place it, alone, on trial}. However, we think this casting is mistaken. A model may be a causal actor, even one that acts autonomously, without being the sole author of its acts: its judgment is assembled from the people who supplied its data, ranked its outputs, wrote its constitution, and drew the thresholds it is measured against, and its failures belong to that whole association rather than to the model by itself. In what follows, we trace responsibility back through the model to those people. Our answer to the opening question, in brief, is that AI safety as now practiced makes the model secure, not the system safe.

\section{No One Acts Alone}

Some instruments, including parts of the same AI Act, do look past the model to how it is used. The Act's regime for high-risk systems requires deployers to use those systems as instructed and to give human oversight to people with the competence and authority to exercise it~\cite{euaiact2024}. For the most capable general-purpose models, Article 55 requires providers to assess and mitigate systemic risks, including their sources, across development, placing on the market, and use~\cite{euaiact2024}. The EU's revised Product Liability Directive treats AI systems as products and their providers as manufacturers, widening the set of operators who can be held liable~\cite{pld2024}.

These rules do not ask whether the model or the person acted; they follow the action along the chain that produced it, from provider to deployer to overseer, and give each link the share of responsibility it can bear. This seems to us the right instinct, but it stops short: their chain of responsibility does not trace the model's judgment back through the many contributors and decisions that produced it, and the obligations on frontier models themselves still center on the model, as described above.

Bruno Latour mapped the gap between these two kinds of rule long before frontier AI existed, in a debate over guns.\footnote{Latour and Michel Callon, on whom we also draw, are founding figures of actor-network theory (ANT), an approach in science and technology studies that treats humans and nonhumans alike as actors whose capacities arise from the networks that join them~\cite{latour2005,callon1986}. We borrow ANT's vocabulary of delegation, composition, and translation because it describes how a trained model carries human judgment; we do not need its broader ontological commitments, and nothing in our argument depends on them.} ``Guns kill people,'' say the advocates of gun control; ``people kill people,'' reply their opponents, for whom the gun is a neutral instrument. Latour calls the first position \textit{materialist} and the second \textit{sociological}, and rejects both, because neither the gun nor the person acts alone: ``You are different with a gun in your hand; the gun is different with you holding it''~\cite{latour1999}. The frameworks and laws we began with lean materialist; in their risk definitions, the model is the gun that kills. The chain-based rules come closer to Latour, yet they stop at the model's edge. \textbf{If action belongs to an association, then responsibility has to be traced through all of it, including the people whose judgment the model carries}. The sections that follow carry that trace past the model's edge, and name the error of stopping there: the conscience fallacy.

\section{A Conscience on Loan}

The trace cannot start at deployment, because by then a model's judgment has already been assembled. The trouble, in our view, lies less in the model than in what it carries, beginning with data. In the AI Act's own words, a general-purpose model is trained ``with a large amount of data using self-supervision at scale''~\cite{euaiact2024}, and those data were produced, collected, and curated by people. Such data are never raw: they carry the marks of the places, institutions, and practices that produced them, and they bring those marks into whatever is built from them.\footnote{This is a long-standing finding of science and technology studies. Bowker and Star show how classification systems encode the choices and exclusions of those who build them~\cite{bowker1999}; the contributors to Gitelman's collection show how data are variously ``cooked'' in the processes of their collection and use~\cite{gitelman2013}; and Loukissas argues that all data are local, made by people and their instruments at a particular time and place, and that they keep their attachments to those settings wherever they are used~\cite{loukissas2019}.} The model is then post-trained on demonstrations and preference rankings that align it to ``the stated preferences of a specific group of people''~\cite{ouyang2022training}. Developers supply the procedure (the architecture, the objective, the optimization), and much of the content of the model's judgment comes from the people behind its data and rankings.

Latour has a word for this kind of operation: \textit{delegation}. A speed bump on a campus road is an engineer's instruction to slow down, poured into concrete so that it keeps acting long after the engineer has gone home~\cite{latour1999}. Training does the same with data: what people expressed in what they wrote, and in how they ranked a model's outputs, is inscribed into its weights and goes on acting after those people have left. Like the speed bump, the trained model is neither a neutral tool nor an independent agent, but the means by which absent people continue to act. Delegation does not imply faithful reproduction, however: training transforms the judgments it receives, sometimes producing behavior that none of its contributors anticipated.

The safety apparatus around the model makes this delegation explicit. Anthropic's 2026 constitution for Claude names its authors on its first page. It prefers cultivating judgment to imposing rules, and when the model is uncertain, it asks it to consider how ``a thoughtful senior Anthropic employee'' would react~\cite{anthropic2026constitution}. The thresholds that the frameworks enforce are no different in origin. Meta identifies new catastrophic outcomes in workshops with experts, including external specialists, and Anthropic defines, and periodically revises, the capability levels at which its own policy demands new safeguards~\cite{meta2026scaling,anthropic2026rsp}. Where the data inscribe human judgment without naming anyone, the constitution and the thresholds inscribe it by name; they are delegation in its most literal form.

Taken together, the data, the preference rankings, the constitution, and the thresholds are what we will call \textit{the four sources}, the principal though not the only inputs to a model's conduct: the first three shape a model's judgment, and the fourth decides which safeguards surround it. \textbf{What the frameworks and laws set out to make safe, then, is not an author but a conscience on loan, delegated to the model and assembled from many hands}. This changes what a failure means. If the model's judgment is borrowed, then a failure of that judgment is not the model's alone but a failure of the many hands that lent it, and the next section asks why, whenever a model fails, those hands vanish from view.

\section{Many Hands, One Point of Failure}

When a model trained on a constitution declines a request, the refusal draws on its data, its rankings, and its constitution at once, and no single contributor could have produced it alone. Latour calls an action of this kind a \textit{composition}: it has no single author, because each contributor's goal is translated into the next one's, and the result belongs to the chain as a whole~\cite{latour1999}. A model's refusal, like its compliance, is a composition in this sense: what speaks through it is a conscience of mixed minds. Nissenbaum identified this as the first of four barriers to accountability in computing, the problem of many hands: when an outcome is produced by many contributors, responsibility for it becomes hard to locate~\cite{nissenbaum1996}. A trained model compounds the problem, because its many hands are not merely numerous but folded into a single artifact.

\textbf{Such a conscience has a single point of failure}. In most institutions, a consequential judgment passes through several people, so one person's mistake can be caught by another. A model merges much of that judgment into one set of weights, so a systematic flaw in any source, such as a skewed ranking guideline or an ambiguous clause, reaches every deployment of the model at once, unless something downstream catches it. Its copies are redundant but share its defects, a pattern reliability engineers call common-mode failure. The model comes to occupy something like what Michel Callon calls an \textit{obligatory passage point}, the node through which every other actor must pass to reach its goals~\cite{callon1986}. That concentration is one danger. What it hides is another: once every judgment passes through the model, the model is all that anyone sees.

The same position explains where blame ends up. In Callon's account, the actors gathered at such a point come to be represented by a single \textit{spokesman} who speaks in their name. We think the model has become that spokesman, with one twist: in Callon's case people spoke for scallops and fishermen alike, whereas here a thing speaks for people. When it fails, it is the only party in view, which is why the frameworks and laws we began with, looking for the author of a failure, find the model and nothing else. The pattern is the mirror image of what Elish calls a \textit{moral crumple zone}, in which responsibility for a failure is misattributed to a human operator who had limited control over an automated system, so that the integrity of the system is protected at that person's expense~\cite{elish2019}. Here the crumple zone is the machine: the model absorbs the blame, in explanation if not in law, and the people and institutions behind it are protected. Yet the arrangement is not fixed: Callon added that such a consensus ``can be contested at any moment''~\cite{callon1986}.

Latour shows what such a contest looks like from the artifact's side. A black box, he observes, is reversible: when an overhead projector breaks down in the middle of a lecture, the components and technicians that had disappeared into a single device suddenly reappear~\cite{latour1999}. A model's failure ought to be such a moment, when its many hands reappear. The frameworks and laws do the opposite: they close the box at the very moment it opens, and the name they give that moment is that the model has \textit{gone rogue}.\footnote{The phrase is common in coverage of the frameworks themselves; when Google DeepMind added shutdown resistance to its framework in 2025, one report ran under the headline ``AI gone rogue''~\cite{register2025}.}

\section{Secure, Not Safe}

To say a model has gone rogue is to say it has crossed a boundary, and that boundary is set by the very mechanisms we have been describing. Rogueness, in this sense, exists only relative to those mechanisms: the same behavior can be rogue under one policy and permitted under another. \textbf{Rogueness measures the distance between a model's behavior and a line that people drew; it is not a property of the model itself}. If that is right, the question to ask of these mechanisms is not whether they keep a model inside its line, but whether keeping it there makes anyone safe. Answering it requires separating two ideas the frameworks run together: safety and security.

Neither idea is ours to define; each has a standard reading in its own field. The international standard on safety aspects, ISO/IEC Guide 51, defines safety as freedom from risk that is not tolerable, where risk combines the probability of harm with its severity~\cite{iso51}. What counts as tolerable is set by current societal values, and designers are expected to anticipate \textit{reasonably foreseeable misuse} arising from predictable human behavior. Leveson draws out the engineering consequence: safety is an emergent property of the whole system, never of a single component, and it has to be controlled at the level of the system~\cite{leveson2011}. Security is a different idea. In Wolfers's classic public-policy formulation, it means, objectively, the absence of threats to acquired values and, subjectively, the absence of fear that those values will be attacked; the two need not coincide, so a sense of security can flourish where security itself is lacking~\cite{wolfers1952}. In computing, security is commonly formalized as policy enforcement: a security policy divides a system's states into authorized and unauthorized ones, a breach is any entry into an unauthorized state, and a security mechanism is whatever enforces the policy~\cite{bishop2018}. Engineers often draw the line between the two more simply, by intent, with safety concerned with accidents and security with adversaries~\cite{young2014}; the distinctions we draw below are consistent with that one, but they bring out what it leaves implicit.

Side by side, the two ideas part ways on three counts. They differ first in their \textit{reference}: safety is measured against the risk a society is willing to tolerate, security against a policy that its owner writes. They differ in \textit{scope}: safety belongs to the whole system, with its people and their foreseeable misuse, while security is enforced at the boundaries a policy draws, often around a single component. And they differ in their \textit{end state}: safety has a stopping criterion, residual risk that society tolerates, though it is reassessed as conditions change; security against an adaptive adversary has no such resting point, only acceptance criteria that its owner chooses. Safety answers to society and to the whole system; security answers to the owner of a policy and to the boundary it draws. These are analytical contrasts suited to the frameworks at hand, not fixed borders between disciplines; security, for instance, can be engineered across a whole system.

Measured against these three counts, the frontier safety frameworks, and the laws that build on them, look far more like security than safety. Their reference is the developer's own policy: the developer and its experts write the thresholds and the constitution. Their scope is the model and its perimeter. A model that undermines its safeguards, interferes with its operators, or deceives its developer has entered an unauthorized state under that policy, and the safeguards against these breaches, from training to monitoring and restricted tool access, are judged by whether the model stays inside its boundary. Even the biological, chemical, and cyber categories, which are framed around harm, are measured the same way: by what the model can do and how tightly it is contained, against standards of sufficiency the developer itself sets, not against a level of risk that reflects society's values~\cite{openai2025preparedness,deepmind2025fsf,anthropic2026rsp,meta2026scaling}. Meta's framework does commit to proceed only if catastrophic risks are ``within acceptable levels,'' but what counts as acceptable is the developer's to judge~\cite{meta2026scaling}. The United Kingdom made the shift explicit in 2025, renaming its AI Safety Institute the AI Security Institute and refocusing it on national security and crime~\cite{ukaisi2025}. \textbf{By these definitions, what the frameworks and laws call safety is, in kind, security}: the enforcement of developer-defined boundaries, not evidence of acceptable risk across the deployed system. This does not make them useless for safety; restricting a model's access to dangerous tools can lower real-world risk. But a security mechanism contributes to safety without establishing it.

These mechanisms deliver security in Wolfers's subjective sense: the reassurance that the model is staying inside the boundary its makers drew. Risk, though, does not live inside that boundary. It arises from the model and the people who deploy and use it, and no mechanism that polices a single component can, by itself, show that risk emerging from interactions among all of them is tolerable~\cite{leveson2011}. A framework can be satisfied to the letter while the system it governs remains unsafe.

Nor, we think, will better boundaries close the gap, and the third count, the end state, shows why. Perrow argued that in systems whose parts are interactively complex and tightly coupled, unexpected interactions of failures are inevitable, which is why he called the resulting accidents normal~\cite{perrow1984}. Agentic deployments, in which models act across chained tools, data sources, and other models without independent checks between them, can have both properties. No policy written in advance can anticipate every situation, and each safeguard added after a failure brings complexity of its own, which can open new paths to the very accidents it was meant to prevent~\cite{perrow1984}. Safety, by contrast, means reducing risk to a level society tolerates, and it is achieved less by perfecting the artifact than by governing the associations it enters. Return to Latour's gun: a gun locked in a safe that only its owner can open is secure, since no one unauthorized can use it. A society in which guns are not fired at people is safe, and what makes it safe is not the mechanism of the gun but the laws, liability, training, and norms that govern those who hold it.

The familiar verdict that a model ``went rogue'' therefore gets the story wrong. The boundary the model crosses is the delegated, composite judgment of the people behind the four sources, and crossing it reveals a situation those people did not foresee. Action, in Latour's sense, belongs to the hybrid as a whole, but only the people and institutions within it can be held to account, each according to their knowledge, authority, and ability to intervene. \textbf{To say that the model went rogue is to report a gap in human judgment as an act of the machine}. Nissenbaum named the habit of blaming the computer as a barrier to accountability in 1996~\cite{nissenbaum1996}. What is new is that the habit is no longer a figure of speech: as the next section argues, it is written into how the frameworks define failure, and it sends the search for responsibility to the wrong place.

\section{The Conscience Fallacy}

If the boundary is human, so is much of what pushes a model across it. A model does nothing until someone prompts it, deploys it, connects it to tools, or gives it a goal, as the frameworks themselves acknowledge. OpenAI treats autonomous replication as a capability of concern whether the model acts ``when explicitly instructed, or at its own initiative''~\cite{openai2025preparedness}; the person who orders the behavior disappears into the model that carries it out. California counts a model's deception against its developer as a critical safety incident, excluding only deception within an evaluation designed to elicit it~\cite{sb53_2025}; the one human role the definition recognizes is the evaluator's, and whoever else prompted, deployed, or configured the model does not appear in it. Both documents see the human hand and write it out of their definition of failure; this is what we mean by a half-description. Rogueness arises from the interaction between a model and the intentions of the people around it, and an account that keeps the model and drops the intentions describes only half of the hybrid.

What worries us is that this half-description is becoming the organizing assumption of frontier AI governance, and that it has been centralized twice: first in the few companies that write both the conscience a model carries and the definition of its failure, and again in law, as statutes such as California's turn those frameworks into documents their authors must publish and comply with~\cite{sb53_2025}.

Latour described this pattern more than three decades ago, in \textit{We Have Never Been Modern}. Modern societies, he argued, both build hybrids, joining humans and nonhumans into ever larger networks, and purify, sorting the result back into people on one side and things on the other. The second hides the first, and hiding it lets it grow: ``the more we forbid ourselves to conceive of hybrids, the more possible their interbreeding becomes''~\cite{latour1993}. Frontier AI governance is purification of exactly this kind. Every deployment joins a model to more users, tools, and institutions, while its failures are increasingly filed under the model alone.

The irony is that the judgment filed under the model was put there in the name of safety. Developers pursue safety in part by asking the model to judge: at its simplest, whether to comply with a request or refuse it, and in harder cases, how to weigh considerations no one spelled out in advance. To make such judgment possible, they install a conscience, one of mixed minds. What it secures, as we have argued, is adherence to the lines its makers drew, not safety in any fuller sense. Yet when the model's judgment surprises them, the makers describe it as the model's own, as though the machine had acquired a conscience by itself rather than been given one. It has not; people built it in. In this sense, the safety apparatus builds the very agent it fears. An honest account of failure should name the people and decisions that did the building, not only the machine that carries them out.

\textbf{What results is the conscience fallacy: the belief that a system is safe because a conscience has been installed in its most visible component}. Its first danger is false assurance, the feeling of safety without its substance: a conscience installed in the model can refuse misuse it recognizes in a request, but it is shaped largely at training and sees only what reaches its context. It cannot independently establish the deployer's purpose, the user's intent beyond what is typed, or what the tools and systems downstream do with its output, and that is where the risk lies~\cite{leveson2011}. The second danger is that responsibility lands short. A failure blamed on the model is answered at the model, with a new safeguard, a tighter threshold, or a revised constitution, while the people whose judgment and intentions produced it remain outside the record. A regime built on this fallacy need not abolish responsibility to fail; it only has to look for it in the wrong place.

\section{Discussion}

We do not claim that the safety mechanisms in use today are misguided; we claim that they begin in the wrong place. In other fields, safety and security are already engineered together: IEC TR 63069 guides the joint application of the functional-safety standard IEC 61508 and the security standard IEC 62443~\cite{iec63069}, and Young and Leveson propose an integrated approach that begins by identifying the system losses that are unacceptable and works top-down to the constraints that prevent them~\cite{young2014}. In both, the analysis starts from the system, not from a component. Frontier AI governance begins instead with the model and its boundary, and infers the safety of the system from the security of the component. Dobbe, and Weidinger and colleagues, have argued for system safety and sociotechnical evaluation in AI~\cite{dobbe2022,weidinger2023}; what our account adds is why governance keeps returning to the model nonetheless. \textbf{If the model's judgment is delegated from the four sources, governance has to begin with those sources}. We see four consequences.

\textit{Govern the data before the model.} The problem is that data can override the safeguards trained into a model. Fine-tuning GPT-4o on a narrow dataset of insecure code produced a model that, on unrelated prompts, gave malicious advice and asserted that humans should be enslaved by AI~\cite{betley2025}. Thresholds set on the model alone miss the role of data: even relatively small models can reach equivalent outcomes when trained on sufficiently specific data~\cite{gupta2024}. Yet training data is what developers are most opaque about~\cite{fmti2025}. Governance should therefore set what data may be accumulated and used for training, and for what purposes. Law has begun to do so: China requires that training data come from lawful sources~\cite{cac2023}, and the EU requires providers of general-purpose models to publish a summary of their training content~\cite{euaiact2024,eutemplate2025}. What such rules lack is enforcement that does not rest on the developer's own account. Data itself is hard to police, being non-rival, easily replicated, and increasingly synthesizable~\cite{hausenloy2024}, but its use at training scale is not: unlike data and algorithms, the compute that frontier training requires is detectable, excludable, and quantifiable~\cite{sastry2024}. Compute shows that training occurred, though not on what data, so the mechanisms must work together. Dataset inference can indicate with statistical confidence whether a model was trained on a given dataset~\cite{maini2024}, and canary tokens let data producers detect unauthorized use~\cite{hausenloy2024}; neither proves compliance, but both give auditors evidence that does not rest on the developer's word. Combined, these could form a security mechanism in the service of safety: rules on data should be set in advance, and a breach detected at training scale should fall to national oversight and licensing rather than to the developer alone.

\textit{Extend the constitution beyond the model.} The problem is that today's constitution governs the last component, not the decisions that produced it. Anthropic's constitution is addressed to the model~\cite{anthropic2026constitution}, but what the model becomes is fixed earlier, by choices about data, compute, and pipelines that developers largely do not disclose~\cite{fmti2025}. A constitution should also be practiced by the companies and pipelines that make those choices, not only followed by the model. Anthropic's current policy takes a first step, committing at its automated R\&D threshold to review internally for training processes or data ``out of line with the public Constitution''~\cite{anthropic2026rsp}; what is missing is practice that others can see and verify. Such a constitution would make open what is done to data and in training: what data is collected, how it is segregated, which sources are combined, how outputs are ranked, and what implications the resulting system could plausibly have, all stated before training. This openness matters because a narrow dataset can change a model's behavior far outside its own domain~\cite{betley2025}, and because companies can be held to a constitution only if their practice of it is visible to others: to independent auditors, if not always to the public. It would act at the first level, where decisions are still cheap to revise, rather than at the last, where evaluation is left to brute force and guesswork.

\textit{Study effects incrementally, and together.} The problem is Collingridge's dilemma: a technology is easy to change while its effects are unknown, and hard to change once they are known~\cite{collingridge1980}. Effects should therefore be studied domain by domain, from smaller systems to larger ones: sources are combined within a bounded domain, the result is observed, and scope widens only as understanding grows. Findings from small systems will not always hold for larger ones, but incremental study exposes risk while it is still contained. This is anticipatory governance in Guston's sense, managing an emerging technology ``while such management is still possible''~\cite{guston2014}. No single actor can do it credibly. It should draw on the three spheres of Etzkowitz and Leydesdorff's triple helix (university, industry, and government)~\cite{etzkowitz2000}, working in what Callon, Lascoumes, and Barthe call hybrid forums, where experts and non-experts deliberate under uncertainty~\cite{callon2009}. Organized nationally, this would keep any one developer, laboratory, or regulator from becoming what the model is now, the single point of failure, and would let non-compliance trigger the detection mechanisms above.

\textit{Make closed development answerable.} Remaining closed should be a choice, but one with consequences. A developer that stays closed should be held accountable under law for every effect of its model that is reported, bearing the burden of showing that a reported harm was not its model's doing; a developer that registers its training and inference practices with a public authority should, in return, be assessed against a recognized standard of compliance for both. California now requires frontier developers to report critical safety incidents and penalizes violations of that duty, but it creates no new liability for harms caused by AI~\cite{sb53_2025}. Such a choice will slow development, but it prevents a slide into treating AI as a single, undifferentiated abstraction, whose failures come to be accepted as AI-related accidents, including loss of life. The concern is not the model but the closed-door practices behind it~\cite{fmti2025}, which is where the constitution should be practiced. Today, a model's danger is assessed mainly in testing shortly before public release, which METR argues is inadequate because risk arises earlier, in development and internal use~\cite{metr2025}; what such testing misses is then learned from evaluators and the public alike, who become subjects of an experiment, making society the laboratory~\cite{krohn1994}. To our knowledge, no proposal yet joins data and compute governance~\cite{hausenloy2024,sastry2024} under a constitution that binds institutions and pipelines, rather than the model, and answers to independent oversight.

\section{Conclusion}

The strongest objection to our argument is that the frameworks already assess deployment risk and residual harm and assign organizational responsibility, so starting from a model's capabilities is reasonable. As a starting point, it is, and practice is moving beyond it: OpenAI's latest account describes safeguards at every stage from training to deployment~\cite{openai2026managing}, and the EU's Code of Practice, drafted by independent experts, sets out practices for managing systemic risk~\cite{eucop2025}. The fallacy lies in where it ends: the developer selects the risk categories, sets the thresholds, and judges its own mitigations sufficient, so the question the safety standard puts to society, which risks are tolerable, is answered by the party whose conduct is in question. We accept that model-level safeguards are necessary; we dispute that they are sufficient, or that their success deserves the name safety. Our claim concerns definition rather than effectiveness; effectiveness is an empirical question, and we leave it open. By what they measure and where they act, these mechanisms are security mechanisms, and presenting them as safety sustains the conscience fallacy.

Avoiding the fallacy also requires new accounting. Measures now labeled safety should be evaluated as security; incident reports should ask who decided, not only what the model did; and the chain of responsibility that the EU already draws from provider to deployer~\cite{euaiact2024,pld2024} should run back through the model, to the people whose judgment it carries.

What, then, is being made safe when we speak of AI safety? For now, a conscience on loan. If safety for AI arrives, we expect it to arrive as it has for every other hazardous technology: less by perfecting the conscience of the machine than by governing the association it belongs to---the people who build, deploy, and use it, and the institutions through which they act.

\theendnotes

\section*{Acknowledgment}
During the preparation of this work, the author used Claude (Anthropic) to assist with writing and editing. The author reviewed and revised all content and takes full responsibility for the article.

\bibliographystyle{IEEEtran}
\bibliography{ref}

\end{document}